\documentclass[twocolumn,preprintnumbers,amsmath,amssymb,prl,floatfix]{revtex4-1}
\usepackage{comment}
\usepackage{graphicx}% Include figure files
\usepackage{dcolumn}% Align table columns on decimal point
\usepackage{bm}% bold math
\usepackage{color}
\newcommand{\red}{\textcolor{red}}

\usepackage{ulem}
\begin{document}

\title{Supersolid Devil's Staircases of Spin-Orbit-Coupled Bosons in Optical Lattices}
%Devil's Staircase Created by Higher Harmonics in Spin-Orbit-Coupled Bose-Hubbard Supersolids
\author{Daisuke Yamamoto$^{1}$}
\email{yamamoto.daisuke21@nihon-u.ac.jp}
\author{Kotaro Bannai$^2$}
\author{Nobuo Furukawa$^2$}
\author{Carlos A. R. S\'a de Melo$^{3}$}
\affiliation{$^1$Department of Physics, College of Humanities and Sciences, Nihon University, Sakurajosui, Setagaya, Tokyo 156-8550, Japan}
\affiliation{$^2$Department of Physics and Mathematics, Aoyama Gakuin University, Sagamihara, Kanagawa 252-5258, Japan}
\affiliation{$^3$School of Physics, Georgia Institute of Technology, Atlanta, Georgia 30332, USA}

\date{\today}% It is always \today, today,
             % but any date may be explicitly specified
\begin{abstract}
We study the emergence of supersolid Devil's staircases of spin-orbit coupled bosons loaded in optical lattices. We consider two- and three-dimensional systems of pseudo-spin-$1/2$ bosons interacting via local spin-dependent interactions. These interactions together with spin-orbit coupling produce length scales that are commensurate to the lattice spacing. This commensurability leads to Devil's staircases of supersolids, with fractal Hausdorff dimensions, which arise from uniform superfluid phases. We show that umklapp processes are essential for the existence of commensurate supersolids, and that without them the Devil's staircase does not exist. Lastly, we emphasize the generality of our results, suggest experiments that can unveil these unusual predictions, and discuss potential applications to the case of $^{87}$Rb.
\end{abstract}
%\pacs{}
\maketitle
{\it Introduction:}
For the solid phase of quantum fluids the 
question - Can solids be superfluids? - has
be asked many years ago~\cite{leggett-1970} 
in the context of solid $^{4}{\rm He}$, where it
was investigated experimentally, but has so far yielded a negative answer~\cite{chan-2012, chan-2013, reppy-2016}. This question essentially provides the origin of the name supersolid: the solid state of a quantum fluid possessing correlated defects that exhibit superfluidity~\cite{chester-1970}. 
The existence of supersolidity is a very important issue with ramifications in the areas of low temperature condensed matter physics $(^4{\rm He})$~\cite{beamish-2007, maris-2012}, astrophysics (neutron star cores)~\cite{rajagopal-2006, samuelsson-2007}, and ultracold atoms and molecules (large spin atoms, dipolar molecules, and spin-orbit coupled systems)~\cite{pfau-2019, modugno-2019, ferlaino-2019, ferlaino-2021, modugno-2021, lewenstein-2002, mitra-2009, ketterle-2017, ho-2011, stringari-2012}.

In condensed matter physics, studies of supersolidity focus on the emergence of superfluid properties
from a solid with crystalline order. 
However, in ultracold atoms and molecules, investigations of 
supersolidity focus on the emergence of crystalline order (solid) from a superfluid. 
Since quantum gases are low density systems, standard solid phases, like those found in 
$^4{\rm He}$, are out of reach. Thus, in the context of ultracold atoms and molecules, 
the reverse question needs to be asked: Can a superfluid be a solid with crystalline order?
This implies that the original definition of supersolidity, described in the opening paragraph, 
needs to be revised to represent a state of matter where both superfluid and solid-like crystalline 
order coexist\red{.} %and are associated with the superfluid order parameter. 
Using the latter concept, some experimental groups have recently reported the existence of 
supersolids in ultracold dipolar bosons with internal
magnetic moments~\cite{ferlaino-2021, modugno-2021}, following earlier experimental indications of 
at least metastable supersolidity~\cite{pfau-2019, modugno-2019, ferlaino-2019}, which had been theoretically 
suggested as a compromise phase between superfluidity and Wigner-crystalization in dipolar Bose gases~\cite{mitra-2009}.
These recent experiments~\cite{pfau-2019, modugno-2019, ferlaino-2019, ferlaino-2021, modugno-2021} have estimulated a flurry of recent theoretical work on supersolid phases of dipolar bosons~\cite{pohl-2019, blakie-2020, recati-2020, stringari-2020, reatto-2021, pfau-2021a, pfau-2021b}. 

While recent experimental and theoretical work about supersolids in ultracold quantum gases focused on 
continuum and trapped (harmonic and boxed) 
dipolar bosons~\cite{pfau-2019, modugno-2019, ferlaino-2019, ferlaino-2021, modugno-2021, pohl-2019, blakie-2020, recati-2020, stringari-2020, reatto-2021, pfau-2021a, pfau-2021b}, 
experimental investigation of supersolidity of dipolar bosons in optical lattices is still lacking, albeit the existence 
of early theoretical work~\cite{lewenstein-2007, sun-2007, danshita-2009, pupillo-2010, yamamoto-2012} describing the existence of insulating, superfluid and supersolid phases.
In contrast to dipolar systems, we propose experiments to create and detect supersolid 
phases of spin-orbit coupled (SOC) bosons loaded in optical lattices, similar to harmonically trapped bosons~\cite{ketterle-2017}. Furthermore, we theoretically study the emergence of supersolid Devil’s staircases {for} %two- and three-dimensional systems of 
pseudo-spin-1/2 bosons via local spin-dependent interactions, like $^{87}$Rb.

{\it Experimental Proposal:} We propose two experimental setups that could be used to create a Devil's staircase of supersolid phases of spin-1/2 bosons with SOC in optical lattices. The simplest case is the creation of either a {two-dimensional (2D)} square or a {three-dimensional (3D)} cubic optical lattice with the application of two counter-propagating Raman beams~\cite{spielman-2011} parallel to the {optical lattice} $xy$ plane, but making angle $\theta$ with respect to the $x$ axis.
The second experimental setup involves the utilization of radio-frequency chips~\cite{spielman-2013} or monolithic
microwave integrated circuits (MMICs)~\cite{sademelo-2020}, where the axis of the spin-dependent momentum transfer can be changed
from the $x$ direction, through a relative rotation of the device with respect to the optical lattice.

{\it Hamiltonian:} To investigate the {supersolid} phases of spin-orbit coupled bosons in optical lattices, we consider the Hamiltonian for a {2D} square lattice
%%%%%%%%%
\begin{eqnarray}
\hat{\mathcal{H}}&=&\sum_{\langle i,j\rangle}\left(\hat{\bm{b}}_i^\dagger \mathcal{T}_{ij}\hat{\bm{b}}_j+{H.c.}\right) +\sum_i\hat{\bm{b}}_i^\dagger\mathcal{M}\hat{\bm{b}}_i\nonumber\\
&&+\sum_{is}\frac{U_{ss}}{2}\hat{n}_{is}(\hat{n}_{is}-1) +U_{\uparrow\downarrow}\sum_i\hat{n}_{i\uparrow}\hat{n}_{i\downarrow},
\label{eqn:hamiltonian}
\end{eqnarray}
%%%%%%%%%
where $\hat{\bm{b}}_i=(\hat{b}_{i\uparrow}~\hat{b}_{i\downarrow})^T$ denotes the annihilation operators of bosons with internal state (pseudo-spin) $s=\uparrow,\downarrow$ at site $i$ of a square optical lattice with lattice vectors $\bm{a}_1=(a,0)$ and $\bm{a}_2=(0,a)$, and $\hat{n}_{is}=\hat{b}_{is}^\dagger\hat{b}_{is}$ counts the local number of $s$ bosons. 
The $2 \times 2$ matrices are $\mathcal{T}_{ij} = -t\exp[- i {\boldsymbol \sigma}_z \bm{k}_T \cdot (\bm{r}_i-\bm{r}_j)]$ and 
$\mathcal{M} = -\mu {\bf 1} + \frac{\hbar\Omega}{2}{\boldsymbol \sigma}_x +\frac{\hbar\delta}{2}{\boldsymbol \sigma}_z$ with
${\boldsymbol \sigma}$ being the Pauli matrices. 
We consider the equal-Rashba-Dresselhaus {SOC}~\cite{spielman-2011} with momentum transfer 
$\bm{ p}_T = \hbar \bm{ k}_T =  
\hbar k_T (\cos\theta \bm{ e}_{x}+\sin\theta \bm{ e}_{y})$ along the direction tilted from the lattice $x$-axis by angle $\theta$ in the $xy$ plane. 
Here, $\bm{ e}_{x}$ ($\bm{ e}_{y}$) denotes the unit vector in the $x$ ($y$) 
direction of the lattice. The Hamiltonian also includes the Rabi 
coupling $\Omega$ and detuning $\delta$, as well as the standard Bose-Hubbard 
parameters: nearest-neighbor hopping $t$, chemical potential $\mu$, and intraspin 
($s=s^\prime$) and interspin ($s\neq s^\prime$) onsite repulsions 
$U_{ss^\prime}>0$ with $U_{\uparrow\downarrow}^2< 
U_{\uparrow\uparrow}U_{\downarrow\downarrow}$ to prevent phase separation. 
We focus on the 2D case, 
but the 3D case is
analogous, in particular, if one uses the same tilt angle 
$\theta$ in the $xy$ plane. 
We explore the model of Eq.~(\ref{eqn:hamiltonian}) in the regime dominated by {$t$} in comparison to 
{$U_{ss^\prime}$}, and investigate the emergence of supersolids from superfluid phases. 

\begin{figure}[t]
\includegraphics[scale=0.36]{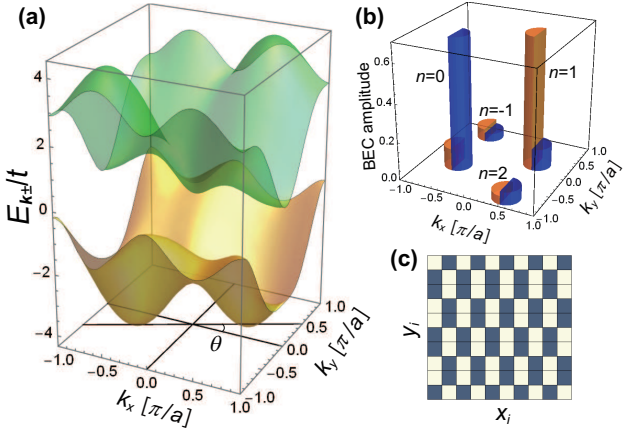}
\caption{
\label{fig1}
(a) Excitation spectra of non-interacting particles $E_{\bm{k}_\pm}$ measured from $\mu$ and (b) the distributions of the BEC components for $\hbar\Omega=3t$, $k_T=0.57\pi/a$, $\theta=0.15\pi$, $\delta=0$,  $U_{\uparrow\uparrow}=U_{\downarrow\downarrow}=U=10t/\rho$, and $U_{\uparrow\downarrow}=0.9 U$. The blue and orange bars in (b) represent the BEC amplitude $|\psi_{n,s}|$ for $s=\uparrow$ and $s=\downarrow$, respectively. 
(c) The stripe pattern of the particle density $\sum_s |\Psi_{is}|^2$, resulting from the multiple BECs of (b), in the real space. The density modulation is $\sim\pm 7.1$ percent of the average filling factor $\rho\gg 1$ for the lattice sites with the lighter and darker colors, respectively.}
\end{figure}
{\it  Multiple Condensates:} Diagonalizing the 
Hamiltonian~Eq.~(\ref{eqn:hamiltonian}) {with $U_{ss^\prime}=0$}, we obtain the excitation spectra of non-interacting particles:
\begin{eqnarray}
E_{\bm{k}\pm}=\frac{\epsilon_{\bm{k}\uparrow}+\epsilon_{\bm{k}\downarrow}}{2}-\mu\pm\sqrt{\left(\frac{\epsilon_{\bm{k}\uparrow}-\epsilon_{\bm{k}\downarrow}}{2}\right)^2+\left(\frac{\hbar\Omega}{2}\right)^2}
\label{spectra}
\end{eqnarray}
with $\epsilon_{\bm{k}s}=-2t[\cos (k_xa+\tau_s{k_Ta\cos\theta})+\cos (k_ya+\tau_s{k_Ta\sin\theta})]+\tau_s{\hbar \delta/2}$ ($\tau_\uparrow=1$ and $\tau_\downarrow=-1$).  
We are interested in the situation where {$\hbar |\Omega|/t$ and $\hbar|\delta|/t$ are sufficiently small to create two minima in the lower branch $E_{\bm{ k}-}$ within the first Brillouin zone (BZ)}. 
For instance, {this is achieved for} $\hbar|\Omega|/t< 4\sum_{j=1,2}|\sin \bm{k}_T\cdot \bm{a}_j\tan \bm{k}_T\cdot \bm{a}_j|$, when $\delta = 0$.
In this case, when the temperature is sufficiently low, the particles form Bose-Einstein condensates (BECs) with wave vectors at the two minima of $E_{\bm{ k}-}$, say, $\bm{ k}_1$ and $\bm{ k}_2$. 
However, in the presence of interactions $U_{ss{^\prime}}$, 
the momenta where condensation occurs are modified to 
$ {\widetilde{\bm{ k}}}_1$ and ${\widetilde{\bm{ k}}}_2$. These momenta can be parametrized as $\widetilde{{\bm{ k}}}_{1} = -\bar{\bm{ q}}+\delta{\bm {q}}$ and $\widetilde{{\bm{ k}}}_{2} =\bar{\bm {q}}+\delta{\bm {q}}$. The
deviation $\delta \bm{ q}$ reflects the parity asymmetry caused by non-zero detuning $\delta$ and/or 
by broken $Z_2$-symmetry interactions 
$U_{\uparrow \uparrow} \ne U_{\downarrow \downarrow}$.

When BECs are formed at $\widetilde{\bm{ k}}_{1}$ and 
$\widetilde{\bm{ k}}_{2}$, the interference of the two fundamental matter waves produces {higher harmonics} with wave vectors differing by an integer multiple of $2\bar{\bm{q}}$. This indicates that the expectation 
value of $\langle\hat{b}_{is}\rangle$ acquires a spatial modulation of the form
\begin{equation}
\label{eqn:multiple-bec}
\langle\hat{b}_{is}\rangle=\Psi_{is}=\sqrt{\rho}\sum_{n} \psi_{ns}e^{i\bm{q}_n\cdot \bm{r}_i}
\end{equation}
with $\bm{q}_n=(2n-1)\bar{\bm{q}}+\delta{\bm{q}}$ {($n\in \mathbb{Z}$)}. Here, $\rho = \sum_{is}|\Psi_{is}|^2/M$ is the particle filling per site, where $M$ is the number of lattice sites. The Fourier amplitudes $\psi_{n s}$ represent BEC order
parameters at each momentum $\bm{ q}_{n}$ and are normalized to one, that is, $\sum_{ns} \vert \psi_{ns} \vert^2 = 1$.
The subscript $n$ labels the harmonic components (HCs), for example, $n=0,1$ correspond to the two fundamental matter waves with wave vectors $\mp \bar{\bm{q}}+\delta{\bm{q}}$, and $n=-1,2$ are the second harmonics with $\mp3\bar{\bm{q}}+\delta{\bm{q}}$, and so on.

Using Eq.~(\ref{eqn:multiple-bec}), we minimize the variational energy per particle
\begin{eqnarray}
\frac{E_0}{M\rho }&=&\sum_{n}\left(\begin{array}{cc}
\psi_{n \uparrow}^\ast  &  \psi_{n \downarrow}^\ast\end{array}\right) \left(\begin{array}{cc}
\epsilon_{\bm{q}_n\uparrow}-\mu &  \hbar\Omega/2\\
\hbar\Omega/2&\epsilon_{\bm{q}_n\downarrow}-\mu\end{array}\right) \left(\begin{array}{c}
\psi_{n \uparrow}  \\  
\psi_{n \downarrow}
\end{array}\right)\nonumber\\
&&+
\sum^\prime_{n_1+n_2=n_3+n_4}\sum_{ss^\prime}
\frac{U_{ss^\prime}\rho}{2} 
\psi_{n_1 s}^\ast\psi_{n_2 s^\prime}^\ast
\psi_{n_3 s^\prime}\psi_{n_4 s},
\label{Ene}
\end{eqnarray}
with respect to the order parameters $ \psi_{ns}$ and the wave vectors $\bar{\bm{q}}$ and $\delta{\bm{q}}$ under the condition $\sum_{ns} |\psi_{ns}|^2=1$.
The sum in the interaction term is over all possible subsets of the HCs that satisfy momentum conservation 
implying the restriction $n_1 + n_2 = n_3 + n_4$. 
When $2\bar{\bm{q}}$ is commensurate to the lattice spacing, we also need to consider umklapp scattering processes with momentum conservation modulo the reciprocal lattice vectors $\bm{G}_1=(2\pi/a,0)$ and $\bm{G}_2=(0,2\pi/a)$, as we shall explain later.

In Fig.~\ref{fig1}(b), we show the ground-state distribution of the order parameter amplitude $|\psi_{ns}|$, in the first BZ, for parameters given in the caption. Remarkably, we find that the $({\bar q}_x, {\bar q}_y)$ components of $\bar{\bm{q}}$, in units of $\pi/a$, may be rational numbers even when the SOC momentum components $(k_{T x}, k_{T y})$, in units of 
$\pi/a$, are irrational 
numbers{, while} the components of {$\delta\bm{q}$}  
can have any real value. For the parameters of Fig.~\ref{fig1}(b), 
where 
{$\delta = 0$} and $U_{\uparrow \uparrow} = U_{\downarrow \downarrow}$, then $\delta \bm{q} = \bm{ 0}$, and the fundamental BEC wave vectors are $\pm\bar{\bm{q}}=\pm(1/2,1/4)\pi/a$. The fundamental wave vectors for the 
density modulation $\sum_s \vert \Psi_{is} \vert^2$ are then 
$ \bm{ Q} = \pm 2\bar{\bm{q}}$, which are expressed as
$\bm{ Q} = \pm \left[ (1/2) \bm{ G}_{1} + (1/4)\bm{ G}_2 \right]$, showing a modulation period {$\Lambda_{x}=2a$ $(\Lambda_{y}=4a)$} along the $x$ $(y)$ direction, as seen in Fig.~\ref{fig1}(c).

When $\bar{q}_x$ and $\bar{q}_y$ are commensurate with $\pi/a$, the number of the HCs is finite. For example, in Fig.~\ref{fig1}(b), the wave vectors of second-harmonic components are given as 
$\bm{q}_{-1,2}=\pm 3\bar{\bm{q}}=\pm(3/2,3/4)\pi/a$, which are equivalent to $\pm(-1/2,3/4)\pi/a$ in the first BZ. In this case, third and higher harmonics are reduced to either the fundamental or the second harmonic wave vectors due to momentum-space periodicity, and thus it is sufficient to consider up to the second components. 
For a general commensurate wave vector $\bar{\bm{q}}=(\xi_1/\eta_1,\xi_2/\eta_2)\pi/a$, with relatively prime integers $\xi_{\ell}$ and $\eta_{\ell}$,
the number of independent HCs is given by 
$N_{\rm HC} = {\rm LCM[\eta_1, \eta_2]}$, where LCM means least common multiple.  
The interference of BECs with $N_{\rm HCs}$ components produces striped interference patterns (supersolids), with wavelengths $\Lambda_{x} = \eta_1 a$ and 
$\Lambda_{y} = \eta_2 a$, in the density profile $|\Psi_{is}|^2$.

{\it Commensurability and umklapp scattering:}
Next, we show that the difference vector $2\bar{\bm{q}}$ between the two fundamental matter waves $\bm{q}_{0,1}=\pm\bar{\bm{q}}+\delta\bm{q}$ must always be of the form $\frac{\xi_1}{\eta_1}\bm{G}_1 + \frac{\xi_2}{\eta_2}\bm{G}_2$ in the ground state, as seen from the conditions satisfied
by the phases $\phi_{ns}$ of the order parameters 
$\psi_{n s}=|\psi_{n s}|e^{i\phi_{n s}}$ for the 
minimization of the ground-state energy in Eq.~\eqref{Ene}. 
First, from the spin-flip terms in Eq.~\eqref{Ene}, the relative phase between $\psi_{n \uparrow}$ and $\psi_{n \downarrow}$ with the same $n$ is determined 
by the sign of $\Omega$ to be $\phi_{n \uparrow} - \phi_{n \downarrow} = \pi$ 
$(\phi_{n \uparrow} - \phi_{n \downarrow} = 0)$ when $\Omega > 0$ ($\Omega<0$). Second, given the previous condition, the interaction terms involving both 
intraspin $U_{ss}$ and interspin $U_{s s^\prime}$ (with $s \ne s^{\prime}$) interactions produce the factor 
\begin{eqnarray}
{\cal A}_{s s^\prime} \cos[\phi_{n_1 s}+\phi_{n_2 s}-\phi_{n_3 s}-\phi_{n_4 s}],
\label{fac}
\end{eqnarray}
with positive interaction coefficients $({\cal A}_{s s^\prime} > 0)$.
Here, we choose the global phase to be $\phi_{0 \uparrow} = -\phi_{1 \uparrow} = -\bar{\phi}/2$, without loss of generality. The second harmonic components $n=2$ and $n=-1$ arise from the scattering processes of the type $(n_1,n_2;n_3,n_4)=(2,0;1,1)$ and $(-1,1;0,0)$, respectively.  Thus, 
their phases must be $\phi_{2 \uparrow}=3\bar{\phi}/2+\pi$ and $\phi_{-1 \uparrow}=-3\bar{\phi}/{2}+\pi$ to minimize the interaction energy in Eq.~\eqref{fac}. 
Analogously, the momentum conservation $n_1 + n_2 = n_3 + n_4$ and the minimization of the interaction energy lead to the conclusion that
the up-spin phases of the HCs satisfy
\begin{equation}
\phi_{n \uparrow} 
=\frac{2n-1}{2}\bar{\phi} + \arccos[(-1)^\frac{|2n-1|-1}{2}]
\label{phasen}
\end{equation}
with $n\in \mathbb{Z}$,
while the down-spin phases are 
$\phi_{n \downarrow} = \phi_{n \uparrow} - \pi$ for $\Omega>0$ or 
$\phi_{n \downarrow} = \phi_{n \uparrow}$ for $\Omega < 0$. 
The last degree of freedom is {the relative phase $\bar{\phi}$ between the two fundamental BECs}, discussed next.

When the difference vector between the two fundamental matter waves is commensurate to the lattice spacing, that is, $2\bar{\bm{q}} = \frac{\xi_1}{\eta_1}\bm{ G}_1 + \frac{\xi_2}{\eta_2}\bm{ G}_2$,
umklapp scattering processes with the total momentum transfer equal to reciprocal lattice vectors 
$
n_1 + n_2 - n_3 - n_4 = \pm{\rm LCM [\eta_1, \eta_2]}
= \pm N_{\rm HC},
$ 
must be considered in the second sum of Eq.~\eqref{Ene}. 
Using Eqs.~\eqref{fac} and \eqref{phasen}, it is easy to see that the sum of the umklapp scattering terms is proportional to {${\cal B} \cos[N_{\rm HC} \bar{\phi}]$}, which shows that the umklapp scattering can always reduce the energy by taking an appropriate value of {$\bar{\phi}$} such that {$\cos[N_{\rm HC}\bar{\phi}] = -1$} $(+1)$ when 
${\cal B} > 0$ $({\cal B} < 0)$. This implies that the supersolid ground state exhibits $N_{\rm HC}$-fold degeneracy with respect to the choice of {$\bar{\phi}$}, and thus breaks $Z_{N}$ symmetry (with $N = N_{\rm HC}$), 
in addition to the breaking ${\rm U(1)}$ symmetry associated with the {global} phase {fixed} earlier.

\begin{figure}[t]
\includegraphics[scale=0.4]{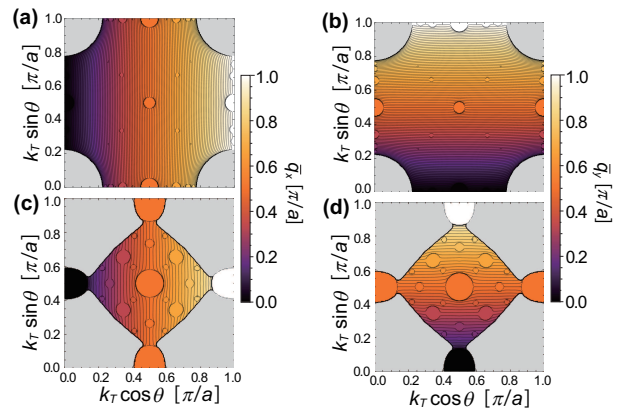}
\caption{\label{fig2} Contour plots of the values of $\bar{q}_x$ and $\bar{q}_y$ in the $k_T \cos \theta$ vs. $k_T \sin \theta$ plane for (a)-(b) $\hbar\Omega=1.2t$ and (c)-(d) $\hbar\Omega=3t$ ($\delta=0$, $U_{\uparrow\uparrow}=U_{\downarrow\downarrow}=U=8t/\rho$, and $U_{\uparrow\downarrow}=0.9 U$ for both). The gray areas represent the regions where a single-BEC superfluid is the ground state.}
\end{figure}
{\it Supersolid Devil's staircase:} 
As discussed above, commensurate ground states with rational values of $\bar{q}_x$ and $\bar{q}_y$ (in units of $\pi/a$) are favored against incommensurate {ones} when umklapp processes are important. This indicates that $\bar{q}_x$ and $\bar{q}_y$ develop plateaus at rational values for numerous intervals of input variables {$k_{T_x}$ and
$k_{T_y}$}, thus forming a Devil's staircase structure with an infinite number of steps. In Fig~\ref{fig2}, we show the values of $\bar{q}_x$ and $\bar{q}_y$ as functions of $k_{T_x} = k_T\cos \theta$ and $k_{T_y} = k_T\sin \theta$ for two sets of parameters {given in the caption.}
{We show only the first quadrant of the first BZ 
since the function $\bar{q}_{x}$ ($\bar{q}_{y}$) is odd in $k_{T_x}$ ($k_{T_y}$) and even in $k_{T_y}$ ($k_{T_x}$).}

\begin{figure}[t]
\includegraphics[scale=0.44]{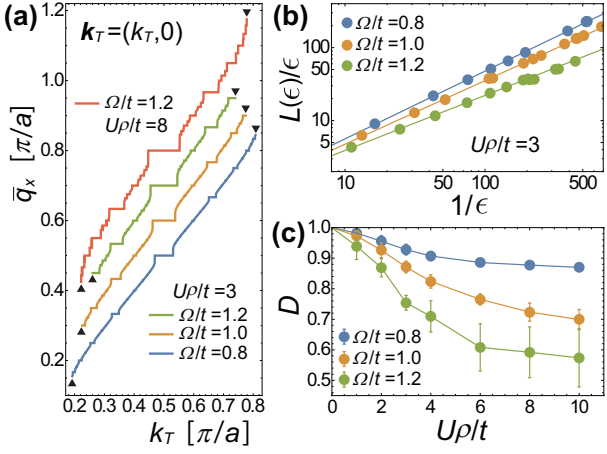}
\caption{\label{fig3} (a) Supersolid Devil's staircases for $\bar{q}_x$ versus $k_T$, in units of $\pi/a$, at SOC angle $\theta=0$. The orange, green, and red lines are vertically shifted by $0.1$, $0.2$, and $0.3$ with respect to the blue line to avoid overlap. 
{The black triangles at the endpoints indicate the emergence of single-BEC superfluid phases.} 
(b) Plots 
of the function $L (\epsilon)/\epsilon$ vs. $1/\epsilon$ characterizing 
the plateau widths $\epsilon$ of the Devil's staircases. (c) Hausdorff fractal dimension $D$ as functions of $U\rho/t$. 
}
\end{figure}

In all panels of Fig.~\ref{fig3}, the parameters used are $U_{\uparrow\uparrow}=U_{\downarrow\downarrow}=U$, $U_{\uparrow\downarrow}=0.9U$, and $\delta=0$. In Fig.~\ref{fig3}(a), we show supersolid Devil's staircases for $\bar{q}_x$ versus $k_T$ at SOC angle $\theta = 0$.  
{A box-counting analysis for the plateau width of the Devil's staircases is shown in Fig.~\ref{fig3}(b)}. The function $L (\epsilon)$ is the difference between the total width of the staircase and the sum of the plateaus widths larger than $\epsilon > 0$. The slope of the log-log plot of $L (\epsilon)/\epsilon$ versus $1/\epsilon$ in the limit of $\epsilon \to 0$ gives the Hausdorff fractal dimension $D$ of the system~\cite{salinas-2014}. If $D < 1$, the incommensurate phases form a fractal set of measure zero, meaning that we have a complete Devil's
staircase of commensurate (supersolid) phases. In Fig.~\ref{fig3}(c), we show $D$ versus $U\rho/t$ to indicate the fractality of the staircase. Notice that when interactions tend to zero ($U\rho/t \to 0$), then the Hausdorff dimension $D \to 1$, meaning that the lines in Fig.~\ref{fig3}(a) do not contain a dense set of plateaus. 
This reinforces that interactions are essential for the emergence of the supersolid Devil's staircase and its fractality.

\begin{figure}[t]
\includegraphics[scale=0.4]{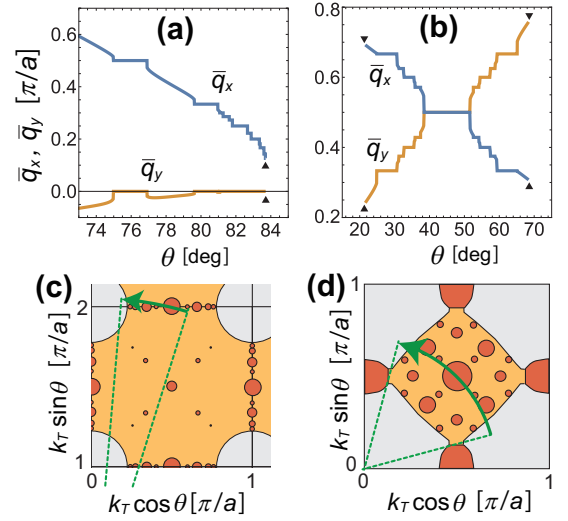}
\caption{
\label{fig4}
Supersolid Devil's staircases for $\bar{q}_{x},\bar{q}_{y}$ vs. SOC angle $\theta$ for fixed values of $k_T$. Panel (a) [(b)] has same parameters as in Figs.~\ref{fig2}(a)-(b) [Figs.~\ref{fig2}(c)-(d)] along the arc with $k_T = 2.02 \pi/a$ $[ k_T = 0.72\pi/a ]$, indicated by the arrow in (c) [(d)]. The triangles have the same meaning as in Fig.~\ref{fig3}(a).
}
\end{figure}

{\it Experimental Detection:} For fixed {$t$, $U_{s s^\prime}$, $\Omega$, and $\delta$}, there are two experimental parameters 
that can be adjusted: 
the magnitude of the momentum transfer $k_T$ and the angle $\theta$ between the direction of {$\bm{ k}_T$} and the $x$ axis of the optical lattice. However, it is easier to vary the tilt angle $\theta$, as changing $k_T$ requires a different laser wavelength for the Raman setup, or a different radio-frequency (microwave) wavelength in the atom-chip (MMIC) configuration. 
Thus, in Fig.~\ref{fig4}, we show examples of {${\bar q}_x$ and 
${\bar q}_y$} for fixed $k_T$ and changing $\theta$. 
In Fig.~\ref{fig4}(a), $k_T = 2.02\pi/a$, but the vector $\bm{ k}_T$
is closer to pointing along the {$k_x$} direction since
$73.5^{\rm o} < \theta < 84.5^{\rm o}$ {[see Fig.~\ref{fig4}(c)]}. {In this case}, the
value of ${\bar q}_y$ is {nearly (or exactly)} zero, while ${\bar q}_x$ take
fractional values with the largest steps being at
${\bar q}_x = \{1/2, 1/3, 1/4, 1/5\}\pi/a$. These are supersolid stripes, where the {density}
$\sum_s \vert \Psi_{is} \vert^2$ is uniform along the $y$ direction and modulated along the $x$ direction with period $\{2, 3, 4, 5\}a$. In Fig.~\ref{fig4}(b), $k_T = 0.72\pi/a$, but the vector $\bm{ k}_T$ is closer to the diagonal {in the first quadrant of the first BZ} [see Fig.~\ref{fig4}(d)]. In this case, 
the most prominent commensurate (supersolid) phases have
the largest ladder steps characterized by the ordered pairs $({\bar q}_x, {\bar q}_y) = 
\{ (1/3,2/3); (2/5, 3/5); (1/2, 1/2); (2/3, 1/3); 
(3/5,2/5) \}$ in units of $\pi/a$. 

The simplest experiments to detect supersolid phases and their staircase structure are momentum space measurements. 
Both the momentum distribution $n(\bm{ k})$ in time of flight~\cite{bloch-2002} 
and the structure factor $S(\bm{q})$ obtained from Bragg spectroscopy~\cite{hulet-2010} can reveal the fundamental 
and higher-order momentum components
of the order parameter density in the supersolid phases.
In addition, the real-space periodic modulations
of the supersolids,
which are commensurate with the underlying lattice structure, can be detected, in principle, using quantum gas microscopes~\cite{greiner-2009, bloch-2016, bakr-2018} 
or magnifiers~\cite{weitenberg-2021}. 

When atoms have anisotropic interactions $U_{\uparrow\uparrow}\neq U_{\downarrow\downarrow}$, then
$Z_{2}$ symmetry is broken. This is the case for two hyperfine states of $^{87}\rm{Rb}$ atoms, $\vert\!\uparrow\rangle = 
\vert F = 1, m_F = 0 \rangle$ and $ \vert\!\downarrow\rangle = \vert F = 1,m_F = -1\rangle$, where $U_{\downarrow\downarrow}\approx U_{\uparrow\downarrow}\approx 0.995 U_{\uparrow\uparrow}$~{\cite{spielman-2011,widera-2006}}. 
The broken spin symmetry can be compensated by the detuning $\delta = \delta_0=-\rho(U_{\uparrow\uparrow}- U_{\downarrow\downarrow})/2$,
since the anisotropy is small $(|U_{\uparrow\uparrow}- U_{\downarrow\downarrow}|\ll \sum_{ss^\prime }U_{ss^\prime})$~\cite{stringari-2012}. In this case, the shift $\delta\bm{q}$ of the BEC momenta is negligible and the results for $U_{\uparrow\uparrow} \ne U_{\downarrow\downarrow}$ with $\delta = \delta_0$ are 
essentially identical to the results for $U_{\uparrow \uparrow}\to {\overline U}$, $U_{\downarrow \downarrow}\to {\overline U}$, and $\delta_0 \to 0$, 
where ${\overline U}=(U_{\uparrow\uparrow}+ U_{\downarrow\downarrow})/2$.

{\it Conclusions:} We have investigated the existence of supersolid Devil's staircases for spin-orbit coupled bosons in optical lattices, proposed  
experimental setups to investigate the phenomenon and suggested detection techniques for their direct observation. Furthermore, we showed that the cascade of supersolid phases occurs due to the commensurability of the spatial modulation of the order parameter with respect to the underlying lattice, which is induced by spin-orbit coupling and stabilized by interactions (umklapp processes). This work opens the door for experimental investigations of supersolidity 
in $^{87}{\rm Rb}$ and other bosonic systems, {and, more generally, provides a fundamental idea for understanding the commensuration of ordering vectors with possible applications to FFLO superconductivity~\cite{Kinnunen-18} and chiral magnets~\cite{tsuruta-16}}.

\begin{acknowledgments}
This work was supported by JSPS KAKENHI Grant Nos.~18K03525 and 21H05185 (D.Y.), JST CREST Grant No.~JPMJCR1673 (D.Y.), and JST PRESTO Grant No.~JPMJPR2118, Japan (D.Y.). 
\end{acknowledgments}

\end{document}